\documentclass[%
 reprint,superscriptaddress,
 amsmath,amssymb,prb,
]{revtex4-2}

\usepackage{graphicx}
\usepackage{dcolumn}
\usepackage{bm}
\usepackage{color}
\usepackage{siunitx}
\usepackage{hyperref}
\usepackage[version=3]{mhchem} 
\usepackage[T1]{fontenc}
\usepackage[utf8]{inputenc}

\usepackage{xcolor}

\begin{document}


\preprint{APS/123-QED}

\title{Magnetism of single crystalline breathing pyrochlore spinel \ce{AgInCr4S8}}

\author{Andrew F. May}
\email{mayaf@ornl.gov}
\affiliation{Materials Science \& Technology Division, Oak Ridge National Laboratory, Oak Ridge, TN 37831, USA}

\author{Christopher M. Pasco}
\affiliation{Materials Science \& Technology Division, Oak Ridge National Laboratory, Oak Ridge, TN 37831, USA}

\author{V. O. Garlea}
\affiliation{Neutron Scattering Division, Oak Ridge National Laboratory, Oak Ridge, Tennessee 37831, USA}

\author{Karolina Gornicka}
\affiliation{Materials Science \& Technology Division, Oak Ridge National Laboratory, Oak Ridge, TN 37831, USA}
\affiliation{Faculty of Applied Physics and Mathematics and Advanced Materials Centre, Gdansk University of Technology, ul. Narutowicza 11/12, 80-233 Gdańsk, Poland}

\author{Matthias D. Frontzek}
\affiliation{Neutron Scattering Division, Oak Ridge National Laboratory, Oak Ridge, Tennessee 37831, USA}

\author{Xiaoping Wang}
\affiliation{Neutron Scattering Division, Oak Ridge National Laboratory, Oak Ridge, Tennessee 37831, USA}

\author{Pyeongjae Park}
\affiliation{Materials Science \& Technology Division, Oak Ridge National Laboratory, Oak Ridge, TN 37831, USA}

\author{Andrew D. Christianson}
\affiliation{Materials Science \& Technology Division, Oak Ridge National Laboratory, Oak Ridge, TN 37831, USA}


\begin{abstract}
Single crystals of \ce{AgInCr4S8} were grown by chemical vapor transport and crystallographic ordering of Ag/In that results in a breathing pyrochlore motif of Cr$^{3+}$ was verified by x-ray and neutron diffraction.  Long-range antiferromagnetic order is observed below a N\'eel temperature of $T_{\mathrm N}$ $\approx$ 9.6 K. The magnetic properties are characterized using ac and dc magnetization, specific heat capacity, and single crystal neutron diffraction measurements.  The specific heat data are characterized by a small lambda anomaly near 9.5 K and the estimated magnetic entropy reaches $\approx$ $\frac{1}{3}$ of the expected value by 3$T_{\mathrm N}$, suggesting significant short-range order in the paramagnetic phase.  Single crystal neutron diffraction evidences an incommensurate spin structure with propagation vector $\textbf{\textit{k}}$ = (0,0,$\delta$) and $\delta$ = 0.343 at 5 K.  The minimal model that accounts for the data consists of ferromagnetic layers of Cr atoms, with magnetic moments lying in the plane of the layers and modulating in the perpendicular direction to form a helical structure propagating along $\textbf{\textit{k}}$.  This study represents a rare investigation of single crystals within the family of breathing pyrochlore materials. This manuscript is published at 10.1103/13q1-pl1s.
\end{abstract}

\maketitle

\section{Introduction}
This manuscript is published at Physical Review Materials \textbf{10}, 054410 (2026); DOI: 10.1103/13q1-pl1s.

The breathing pyrochlore lattice consists of tetrahedrons that share corners and alternate in size.  Chromium forms a breathing pyrochlore sublattice in quaternary $A^{1+}A^{3+}$Cr$_4$(O,S,Se)$_8$ materials that contain an ordered arrangement of $A^{1+}$ (Li,Cu,Ag) and $A^{3+}$ (Al,Ga,In) cations.\cite{Okamoto2013breathing,talanov2020formation}  The formation of this non-centrosymmetric lattice is driven by the coupled size/valence differences in the $A$-site cations, and chromium-containing materials are unique among transition metals in their ability to form such ordered variations.\cite{Mazzotti2025} Breathing pyrochlore spinels are of interest for their diversity of magnetic properties caused by competing interactions.\cite{Okamoto2013breathing,okamoto2018magnetic,Hallas2021,Nilsen2025}  These materials have long-range interactions that help stabilize complex magnetic orders and interesting correlated states due to the presence of degenerate magnetic ground states.\cite{ghosh2019breathing,Pokharel2020Cluster}  Due to this frustration, spin-glass like ground states are prominent.\cite{macikazek2007magnetic,Duda2008} Like their ternary spinel relatives, the breathing pyrochlore spinels often possess strong spin-lattice coupling,\cite{Gao2024dynamic} which can be manifested as a negative thermal expansion\cite{Pokharel2018} or a frustration-relieving lattice distortion.\cite{Nilsen2015,gen2024spin} 
In \ce{CuGaCr4S8} and \ce{CuAlCr4S8}, which distort at their magnetic transitions,  $\frac{1}{2}$ magnetization plateaus are observed and recent work has shown that the cubic phase is recovered at the magnetization plateau in \ce{CuGaCr4S8} above 400 kOe (40 T).\cite{Gen2025}  These interesting examples of the spin-lattice coupling in quaternary spinels represent a small fraction of the behavior observed in the much broader family of spinels, which are generally known for their complex magneto-structural behaviors.

While the magnetic properties of these materials remain of great interest, few examples of breathing pyrochlores materials that can be synthesized in single crystal form are known and detailed characterizations of such samples remain limited.\cite{pinch1970some,macikazek2007magnetic,Duda2008,ghosh2019breathing,Hallas2021} In 1970, Pinch $\textit{et al}$ reported that small single crystals of \ce{AgInCr4S8}, \ce{CuInCr4Se8}, \ce{CuGaCr4Se8}, and \ce{AgInCr4Se8} can be grown using vapor transport with \ce{CrCl3} as a transport agent.\cite{pinch1970some} We note that details for the growth of ternary spinels and some related alloys are summarized by H. von Philipsborn.\cite{Philipsborn1971crystal}   Here we are interested in \ce{AgInCr4S8}, which is listed as antiferromagnetic with a N\'eel temperature ($T_{\mathrm N}$) of 14 K,\cite{pinch1970some} though the corresponding data were not shown.  Neutron diffraction experiments were later performed on a polycrystalline sample of \ce{AgInCr4S8} and an incommensurate, helimagnetic ground state was reported below $T_{\mathrm N}$ = 17 K.\cite{plumier1971mise}  The data in those studies did not evidence the ordering of Ag/In and thus the crystal structure was believed to be a disordered spinel with space group $Fd\bar{3}m$ (No. 227).  However, infrared spectroscopy later evidenced a mode associated with \textit{A}-site order, suggesting at least partial order.\cite{haeuseler1977}  Our study is motivated by the desire to understand the magnetism in \ce{AgInCr4S8} while confirming the nature of the underlying crystal structure.  Neutron diffraction on single crystals is an ideal method for this study due to its ability to differentiate between Ag/In and probe the magnetic order.

In this article, we report the growth and physical properties of \ce{AgInCr4S8} single crystals.  The crystals were grown in the presence of excess sulfur and diffraction data evidence the ordered breathing pyrochlore crystal structure (space group No. 216).  We use simple steric quantities derived from ionic radii to discuss what factors appear to determine whether related sulfides form in the ordered breathing pyrochlore lattice. Our measurements evidence that an antiferromagnetic ground state with long-range order occurs below approximately 10 K in these crystals.  Significant short-range correlations exist above the N\'eel temperature, consistent with the expectation of competing magnetic interactions.  Below $T_{\mathrm N}$, a subtle meta-magnetic transition is observed in isothermal magnetization measurements and saturation of the moment approaching the expected 3$\mu_B$/Cr is achieved by roughly 25 kOe.  The magnetic structure was investigated by single crystal neutron diffraction and an incommensurate spin structure with propagation vector $\textit{k}$ = (0, 0, $\delta$) is observed below $T_{\mathrm N}$ with $\delta$ = 0.343 at $T$ = 5 K.  We present the simplest solution that fits the neutron diffraction data, which is a simple helix rotating while propagating along $\textit{k}$.

\section{Experimental details}

\subsection{Crystal Growth}

Single crystalline \ce{AgInCr4S8} was grown by chlorine-assisted vapor transport starting with polycrystalline source material.  Pinch \textit{et al} noted the ability to grow small crystals of \ce{AgInCr4S8} and a few related quaternary compounds using chromium chloride as the vapor transport agent and this inspired our pursuits.\cite{pinch1970some}  First, \ce{AgInCr4S8} powder was synthesized from Ag (premion shot 5N), In (puratronic shot 5N), Cr (Alfa Aesar powder 99.97\%) and S (puratronic lumps) that were loaded in stoichiometric ratio with an additional 20 mg of sulfur (approx. 5 g total batch size).  The reaction occurred in a fused silica ampoule that was sealed under vacuum after argon purging.  The ampoule was heated in a box furnace in a step-wise manner to avoid catastrophic tube failures, as follows: ramp at 30 $^{\circ}$C/h to 450$^{\circ}$C, dwell for 12 h; ramp at 20 $^{\circ}$C/h to 600 $^{\circ}$C, dwell for 24 h; ramp at 10 $^{\circ}$C/h to 800 $^{\circ}$C, dwell for 24 h; ramp at 20 $^{\circ}$C/h to 1000 $^{\circ}$C, dwell for 48 h; cool to 800 $^{\circ}$C at 100 $^{\circ}$C/h, dwell for 48 h; lastly, turn off the furnace and allow sample to cool in the furnace.

Single crystals were grown by combining 0.5 g of the \ce{AgInCr4S8} powder, 0.1 g of \ce{CrCl3} as a transport agent, and 36 mg of sulfur representing approximately 2 atmospheres of sulfur pressure at temperature; the elements were sealed under vacuum in a fused silica tube.  The ampule was sealed to a length of approximately 24 cm and placed in a three-zone furnace with a separation between zones of approximately 20 cm; one end of the tube lined up with the first zone and the other end towards zone 2.  Temperatures set points for the zones were 805 $^{\circ}$C (zone 1), 760 $^{\circ}$C (zone 2) and 770 $^{\circ}$C (zone 3) for a duration of 75 days.  The higher temperature of zone 3 relative to zone 2 discouraged crystal growth at the very end of the tubes, and crystals grew  along the latter $\approx$ 14 cm of the tube.  At the end of the growth, zone 1 was first cooled to condense the transport agent and sulfur away from the crystals as the furnace cooled.  The tubes were opened and crystals were mechanically removed from tube walls, briefly sonicated in water, then washed in acetone, methanol, and isopropyl alcohol in sequence to clean and dry them.  The crystals were found to be stable in air.  

Many growths were performed and a few key observations regarding the use of excess sulfur are shared here.  Firstly, adding excess sulfur prior to the crystal growth appears to reduce the size of the crystals grown by vapor transport, though we did not optimize \ce{CrCl3} after optimizing sulfur content.  For identical conditions, the largest crystal grown under excess sulfur was 6.5 mg, while the largest crystal grown without excess sulfur was 16 mg. However, the use of excess sulfur was correlated with a sharper anomaly in the specific heat  that is more consistent with long-range magnetic order, and the transition was shifted to higher temperatures.  This suggests sulfur vacancies, and perhaps compensating disorder/non-stoichiometry of Ag/In, are promoted when the material is grown without excess sulfur.  While the physical properties clearly indicated a subtle change based on the growth environment, we were unable to detect a structural difference using x-ray diffraction; the concentration of vacancies is expected to be small.  A similar effect may exist in the polycrystalline samples of this phase and related materials, rendering synthesis under excess chalcogen vapor pressure a useful tool when developing these materials.  Regarding crystal size, we found that excess sulfur beyond an estimated 2 atm further suppresses crystal growth, with excess sulfur between 0.5 and 2 atm resulting in a similar suppression of growth relative to the growths without excess sulfur.  Finally, it was noticed that using \ce{CrCl3} stored under ambient conditions produced larger crystals than \ce{CrCl3} stored in an inert atmosphere glovebox, suggesting that small amounts of adsorbed water may positively impact this growth.  Understanding the mechanisms dictating how growth conditions impact the properties of these materials would likely be a fruitful endeavor that is beyond the scope of our current effort.

\subsection{Characterization}

Sample purity and crystal orientations were initially checked using a powder x-ray diffractometer (PANalytical X’Pert Pro MPD) with Cu--K$_{\alpha1}$ radiation ($\lambda$ = 1.5406\,\AA) from an incident beam monochromator.  The crystal quality was inspected during single crystal x-ray and neutron diffraction experiments and these measurements revealed only the expected diffraction with intensities well-fitted to the crystal structure model.  The single crystal x-ray diffraction was performed using Mo K$_{\alpha}$ 0.71073\,\AA. The data were solved using SIR2019 and refined within WinGX using ShelX.\cite{farrugia2012wingx,sheldrick2008short}  The structure was standardized using Structure Tidy within Platon.\cite{platon}  A Hitachi TM3000 electron microscope equipped with an Bruker Quantax 70 energy dispersive spectroscopy (EDS) system was utilized to inspect the crystals and characterize the relative Ag/In ratio; several crystals were examined and the average result with error bars being the standard deviation is reported.

Physical property measurements were performed in Quantum Design measurement systems. The specific heat capacity was measured in a Physical Property Measurement System (PPMS) employing the relaxation time method with both small and large thermal pulses to obtain sufficient data coverage.  The ac magnetic susceptibility was measured using the ACMS-II option in a PPMS; the ac driving amplitude was 3 Oe and a range of frequencies were utilized.  The dc magnetization data were collected in a Quantum Design SQUID magnetometer (MPMS3) using dc measurement scans.  DC magnetization data were collected in both the zero field cooled (ZFC) warming and field cooled (FC) cooling protocols.

Neutron diffraction experiments were performed on a crystal of mass 6.47 mg using the WAND$^2$ diffractometer \cite{frontzek_2018} at the High Flux Isotope Reactor with wavelength, $\lambda$ = 1.486 \AA. The crystal was aligned with the $(H,K,H)$ scattering plane approximately horizontal. The sample temperature was controlled by a variable temperature insert with a base temperature of 1.5 K. Data were collected by rotating the sample about its vertical axis through an angular range of 125$^{\circ}$ with a step size of 0.1$^{\circ}$ and a counting time of 40 sec per angle. The data were reduced using Mantid Workbench \cite{mantid_2014}. The neutron diffraction data were analyzed using the FullProf Suite software package \cite{fullprof}.  The magnetization of the crystal utilized was measured prior to neutron diffraction and the behavior is consistent with that reported here.

\section{Results \& Discussion}

\subsection{Crystal Structure and Atomic Ordering}

$A$-site cation ordering (e.g. Ag$^{1+}$/In$^{3+}$) leads to a lattice distortion in the  breathing pyrochlore spinels relative to quaternaries with $A$-site disorder or ternaries with a single divalent cation (e.g. $A$ = Cd in \ce{CdCr2S4}). The reduction in symmetry is associated with a loss of the inversion center and a change from space group No. 227 ($Fd\bar{3}m$) to No. 216 ($F\bar{4}$3$m$).  Previously, x-ray diffraction had not clearly identified the $A$-site order in \ce{AgInCr4S8} but infra-red spectroscopy suggested $A$-site ordering.\cite{haeuseler1977}  We note that Ag has a larger coherent scattering length (5.922 fm) compared to In (4.065 fm), which leads to measurable contrast in neutron diffraction measurements, though modern single crystal x-ray diffraction instrumentation was also found to be sensitive enough to observe the reduction in symmetry.    Cation ordering can be evidenced by the observation of Bragg reflections that are prohibited in space group No. 227.  Specifically, we observed the 200, 420 and 600 Bragg reflections that are allowed in $F\bar{4}3m$ but forbidden in $Fd\bar{3}m$.  Thus, we confirmed the \textit{A-site} ordering and associated breathing pyrochlore structure in \ce{AgInCr4S8} via diffraction measurements.  In addition, the single crystal diffraction experiments illustrated a good quality of the crystal in that we did not observed additional scattering and the peaks were relatively sharp and intense with intensities well fitted by the breathing pyrochlore structure.

\begin{figure}
\includegraphics[width=0.96\columnwidth]{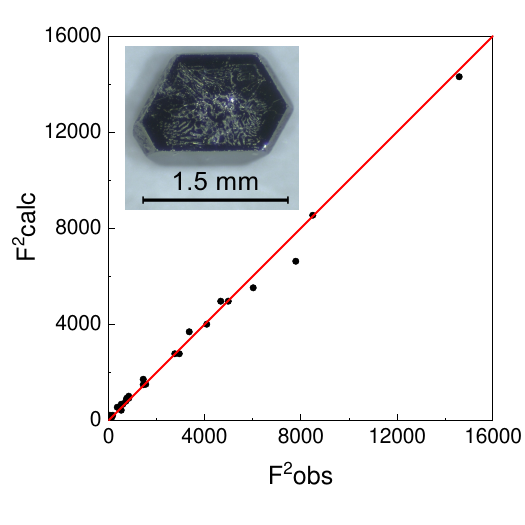} 
\caption{\label{nuclear} Fitted nuclear intensities from neutron diffraction data collected at $T$ = 5 K for \ce{AgInCr4S8}. Refinement assuming a fully ordered structure gives R$_f$-factor = 3.9\%  on 25 reflections. A picture of a typical single crystal is shown as an inset.}
\end{figure}

A comparison of the computed and observed neutron diffraction intensity for the nuclear structure is shown in Fig. \ref{nuclear}.  Our refinement indicates complete occupancy of In on its position, suggesting a well-ordered crystal structure. However, up to $\sim$ 20\% In was refined on the Ag position when constraints were not in place, though the quality factor did not improve. Such a large concentration of anti-site defects would result in a significant change from the anticipated stoichiometry, which we did not detect.  Specifically, our energy dispersive spectroscopy data reveal a nearly equal concentration of In and Ag in the crystal, within the error bars (absolute ratio In/Ag = 1.02(2) where the standard deviation obtained by averaging many points is given in parenthesis).  Thus, with the observation of forbidden peaks and the corresponding refinement,  we believe Ag/In are well-ordered in the crystals reported upon in this manuscript and that the actual composition is close to \ce{AgInCr4S8}.

\begin{table}[!h]
  \caption{Data from refinements of single crystal x-ray diffraction for \ce{AgInCr4S8} at ambient condition. All atoms reside at ($x$,$x$,$x$) with $x$ provided. See the supplemental materials for .cif file.\cite{supp}}
  \label{tab_scxrd}
  \begin{tabular}{lc}
   \hline
    \\
   Space Group                 &   $F\bar{4}$3$m$ (216) \\
   \textit{a}  (\AA)           &   10.20020(10)         \\
   vol (\AA$^3$)               &   1061.27(3)          \\
   density (g/cm$^3$)          &   4.301        \\
   Ag: $x$, U$_{\textrm{iso}}$        &  0, 0.0136(3)     \\
   In: $x$,  U$_{\textrm{iso}}$       &  $\frac{3}{4}$, 0.0080(2)   \\
   S1: $x$, U$_{\textrm{iso}}$        &  0.13906(13), 0.0080(3)    \\
   S2: $x$, U$_{\textrm{iso}}$        &  0.61155(13), 0.0079(4)   \\
   Cr: $x$,  U$_{\textrm{iso}}$       &  0.37104(10), 0.0066(2)   \\
   R$_1$  (all data)           &   0.0157    \\
   wR$_2$ (all data)           &   0.0397         \\
   R$_{\mathrm{int}}$          &   0.0257      \\
   Goodness of Fit			    & 	1.351	        \\
   highest peak/deepest hole   &   2.6/-4.65 \\
   data/params	                &   281/10      \\
   reflections/rejected		&   3931/0   \\
   \hline
  \end{tabular}
\end{table}

In Table \ref{tab_scxrd}, we report structural parameters from the refinement of single crystal x-ray diffraction data for \ce{AgInCr4S8}.  In addition to $A$-site ordering, the symmetry reduction allows Cr to move off the special position and reside at (x,x,x).  This leads to the geometric breathing ratio, $B_r$, defined using the ratio of the Cr-Cr distance within the two distinct \ce{Cr4} tetrahedra. For these quaternary spinels, $B_r$ = (1-2$x$)/(2$x$-0.5) for $x$ $<$ 0.375.\cite{Pokharel2018}  In \ce{AgInCr4S8}, $B_r$ = 1.065 is obtained based on the room temperature structural parameters reported in Table \ref{tab_scxrd}. This is similar to the value of 1.070 reported for \ce{LiGaCr4S8}\cite{Pokharel2018}. Since this is a distance ratio, it does not depend on the unit cell volume (lattice constants). 

A distortion from perfect octahedral coordination (CrS$_6$) occurs in these quaternary sulfides and thus the magnetic interaction via Cr-S-Cr are similar though variable (Cr-Cr distances are well over 3 \AA\ and direct exchange is probably not dominant).  Specifically, the crystal structure enforces two different Cr-S-Cr exchange paths with bond angles that are always slightly above 90 degree.  For those with modern crystallographic data, the largest variation between the two Cr-S-Cr bond angle occurs in \ce{CuInCr4S8}, then \ce{LiGaCr4S8} and the least variation in \ce{AgInCr4S8}; the ratios of bond angles are as follows: \ce{CuInCr4S8}:1.078, \ce{LiGaCr4S8}: 1.047, \ce{AgInCr4S8}: 1.032.  This structural change may impact the relative values of the nearest-neighbor exchanges within the small and large tetrahedron, J and J', respectively.  However, it appears that J,J' are ferromagnetic for the sulfides; the second neighbor interaction is generally weak but a dominant set of non-equivalent third-neighbor antiferromagnetic interactions (J$_{3a,b}$) cause magnetic frustration.\cite{ghosh2019breathing}

We now discuss our observations regarding the formation of breathing pyrochlore spinels that contain sulfur as the anion.  The necessary \textit{A}-site ordering is linked to the intertwined size/charge variation between the two cations.  We first emphasize that all relevant materials (ordered/disordered variants) have the same valence difference $A^{1+}$ and $A^{3+}$.  Therefore, ionic radii are likely to play a key role in stabilizing the ordered lattice.  We note that Pinch \textit{et al} concluded that \textit{A}-site ordering was only present when the quaternary contained either Li or Cu and was not impacted by the trivalent cation.\cite{pinch1970some}  This conclusion was based on the available data, which failed to provide evidence for \textit{A}-site ordering for any of the Ag-based compounds they investigated.  However, from our results and prior optical data, we now know that \ce{AgInCr4S8} also has $A$-site order.  We have thus reconsidered what drives the formation of the breathing pyrochlore lattice in these materials in hopes of identifying other candidate materials.

\begin{table}[]
\caption{Table of experimentally observed ordering and associated steric parameters in $A^{1+}A^{3+}$\ce{Cr4S8} spinels; values computed using Shannon radii\cite{shannon1976revised}, including the spinel tolerance factor $t$ from \citenum{song2020tolerance}; a radius of 1.84\AA was used for divalent sulfur.}
\label{tab_order}
\begin{tabular}{|c|c|c|c|c|c|}
\hline
$A^{1+}A^{3+}$ & $A$-site order & $R^{1+}$/$R^{3+}$ & $R^{1+}$-$R^{3+}$ & $R_{ave}$ & $t$ \\ \hline
LiIn   & yes          & 0.95      & -0.03           & 0.61   & 0.87               \\ \hline
CuIn   & yes          & 0.97      & -0.02           & 0.61   & 0.87               \\ \hline
LiGa   & yes          & 1.26      & 0.12            & 0.53   & 0.90               \\ \hline
CuGa   & yes          & 1.28      & 0.13            & 0.54   & 0.90               \\ \hline
CuAl   & yes          & 1.54      & 0.21            & 0.50   & 0.91               \\ \hline
AgIn   & yes          & 1.61      & 0.4             & 0.81   & 0.80               \\ \hline
AgGa   & no           & 2.13      & 0.55            & 0.74   & 0.83               \\ \hline
AgAl   & no           & 2.56      & 0.63            & 0.70   & 0.84               \\ \hline
\end{tabular}
\end{table}

We considered several steric factors to identify one that provided a clear demarcation between ordered/disordered variants.  As shown in Table \ref{tab_order}, we calculated the radius ratio $R^{1+}$ / $R^{3+}$, the difference $R^{1+}$ - $R^{3+}$, the average radius, and the spinel tolerance factor established by Song and Liu.\cite{song2020tolerance}.  From these, only the radius ratio (or absolute difference) provides a singular division between materials with A-site order/disorder.  Ordering is observed when $R^{1+}$ / $R^{3+}$ is less than or equal to $\approx$ 1.61, or the absolute difference is less than $\approx$ 0.4, with this upper bound being observd in the title compound \ce{AgInCr4S8}.  In this formulation, \ce{AgGaCr4S8} and \ce{AgAlCr4S8} have a radius ratio (or difference) that is too large for chemical ordering to occur.   This observation is limited by the relatively few examples but seems worth pointing out.  There are no examples with a radius ratio slightly larger than that found in \ce{AgInCr4S8} so we do not know if ordering would exist for a ratio of 1.8, for instance.

It may seem counterintuitive that a larger size difference promotes disorder; indeed, the size-difference is often a fundamental driving force for atomic ordering.  The breathing pyrochlore spinels seems to be a somewhat unique case.  For instance, ordering exists in \ce{CuInCr4S8} despite the negligible radii difference between Cu$^{1+}$ (0.6\AA) and In$^{3+}$ (0.62 \AA).   This indicates that valence/charge is a dominant factor in breathing pyrochlore spinels.  We therefore presume that the internal lattice strain destabilizes the ordered structure when the size difference becomes very large, in which case the entropic term dominates and disordered cubic spinels form.  The fact that these materials still form a cubic spinel structures may be a surprise, but their spinel tolerance factor is found to be within the range formed by the breathing pyrochlore spinels.  We utilized the average of the $A$-site radii when computing the spinel tolerance factor.

We do not easily identify hypothetical or unexplored compounds using the radius ratio criterion observed here.  The guiding principle would be to find cation combinations with stable valence states and reasonable radii.  Na and Ag have very similar ionic radii and thus analogous the ordering criteria for hypothetical Na-based breathing pyrochlores would track that of the Ag-based ones, though we are not aware of any Na-based quaternary spinels.  

Considering the trivalent cations: Sc$^{3+}$ has an appropriate radii for hypothetical compound \ce{AgScCr4S8} but Sc tends to occupy the \textit{B}-site in related spinels (e.g. \ce{CdSc2S4}\cite{pawlak1981influence}).  Surprisingly, \ce{CuFeCr4S8} has been reported to form in the breathing pyrochlore lattice with Fe$^{3+}$ on an A-site,\cite{sadykov2001neutron} and the radius ratio (1.22) falls in the middle of the stability range for A-site order; hypothetical \ce{AgFeCr4S8} has a radius ratio of 2.08 and would be expected to form a disordered variant if the quaternary forms.   

Despite the example of \ce{CuFeCr4S8}, finding a trivalent cation with a filled \textit{d}-shell may be important, with Sb and Tl being obvious candidates.  Sb has been reported to substitute for Cr in \ce{CuCr2S4}\cite{krok2004influence}, which is a bit surprising since In$^{3+}$ does not appear to substitute for Cr$^{3+}$ in \ce{AgInCr4S8} even though it is closer in size to Cr$^{3+}$ than is Sb$^{3+}$. It should be noted that In is the largest trivalent cation to support the breathing pyrochlore spinel lattice, and thus Sb may simply be too large (leading to a large negative size difference $R^{1+}$-$R^{3+}$).  This raises the question of a lower limit to the radius ratio when much less than unity, with $A$-$B$ site mixing possibly becoming dominant. A very interesting realization based on Tl incorporation would be a breathing pyrochlore compound of the form Tl$^{1+}$Tl$^{3+}$Cr$_4$S$_8$.  Even if the mixed-valence behavior were energetically favorable, the stereoactive lone pair of Tl may hinder the formation of such a phase by disturbing the local symmetry.  

Some of these hypothetical compounds might be metastable and could be formed via low temperature synthesis routes that minimize the role of entropy.  Computational studies of the factors dictating stability, including the potential for anti-site defects in some of the hypothetical compounds mentioned above,  could facilitate the discovery of new breathing pyrochlore spinels.

\subsection{Magnetization and Specific Heat}

\begin{figure*}[ht]
\includegraphics[width=1.96\columnwidth]{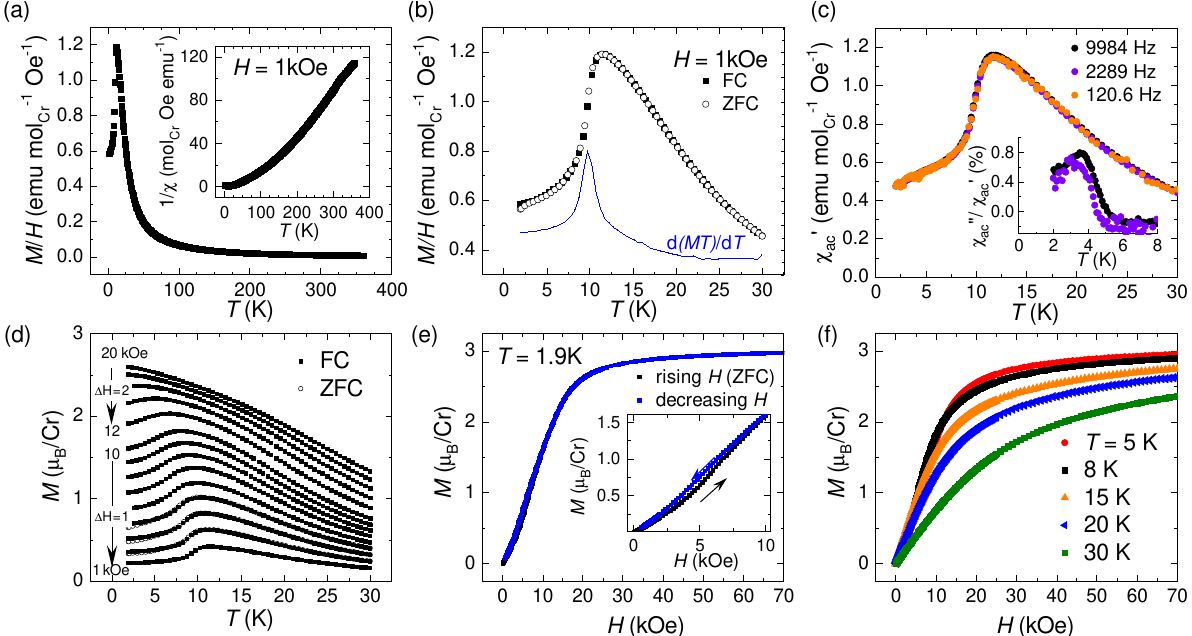} 
\caption{\label{mag} Magnetization data for $H$ $\parallel$ [111]. (a) Magnetization versus temperature with inset showing the inverse susceptibility.  (b) Magnetization near the magnetic ordering transition with derivative d$(MT)$/d$T$ shown as a thin line that has a peak near 9.7 K.  (c) The in-phase component of the ac magnetic susceptibility $\chi'$ for various frequencies as provided in the legend, with inset showing a negligible out-of-phase $\chi''$ observed until a small $\chi''$ forms near 3.8 K.  (d) Temperature-dependent magnetization for various applied fields as indicated. (e) Isothermal magnetization at 1.9 K with inset emphasizing a small region of hysteresis, and (f) isothermal magnetization at various temperatures as indicated.}
\end{figure*}

We report magnetization ($M$) data for measurements with the magnetic field $H$ applied along the [111] crystallographic direction in Fig. \ref{mag}.  The magnetization increases on cooling as expected for a local-moment paramagnet, down to approximately 12 K.  This behavior is illustrated in Fig.\ref{mag}(a), where the inset presents the data as the inverse susceptibility (1/$\chi$ = $H$/$M$).  The $1/\chi$ data possess a non-linear temperature dependence, demonstrating that fitting to Curie-Weiss behavior is not necessarily appropriate.  Upon considering all the data presented below, we believe that the deviation from Curie-Weiss behavior is due to the existence of persistent short-range correlations among spins at high temperatures, a phenomenon commonly observed in breathing pyrochlore spinels and other frustrated magnetic systems.  However, to provide additional data for consideration, we fit the $\chi$ data between 360 and 150\,K to a modified Curie-Weiss law ($\chi$ = $\chi_0$ + $C/(T-\Theta)$, where $\Theta$ is the Weiss temperature and the Curie constant $C$ is related to the effective moment $\mu_{eff}$).  This fit produced $\Theta$ = 56.8(7)K, confirming the strong ferromagnetic interactions, as well as an effective moment of 5.4(5)$\mu_B$/Cr and $\chi_0$ = -0.0038(1) emu/Oe/mol-Cr.  To determine the influence of $\chi_0$, we fixed $\chi_0$=0 and fitting to the same range produces $\Theta$ = 79.1(2)K and  $\mu_{eff}$ = 4.56(2) $\mu_B$/Cr.  In both results, the effective moment is larger than the 3.87$\mu_B$/Cr expected for spin-only Cr$^{3+}$, indicating a possible orbital contribution or error resulting from the use of an inappropriate fitting due to the presence of short-range correlations.”

The dc magnetization in the region of the N\'eel temperature is examined in Fig.\ref{mag}(b) and comparative data for the ac susceptibility are shown in Fig. \ref{mag}(c).  Slightly above $T_{\mathrm N}$, $M$ displays a broad peak and a sharp decrease in $M$ is observed upon cooling.  A sharp peak in d$(MT)$/d$T$ is observed near 9.7 K and this can be utilized to define a N\'eel temperature;\cite{fisher1962relation} a peak occurs at a nearly identical temperature in d$M$/d$T$.  The ac and dc susceptibility have very similar temperature dependence and magnitudes.  As shown in Fig. \ref{mag}c, we did not observe any frequency dependence to the susceptibility in the vicinity of the N\'eel temperature for frequencies ranging from 120 to 9998 Hz.  As such, the out-of-phase component $\chi''$ is negligible.  However, a small onset of $\chi''$ is observed at the lowest temperatures, as shown in the inset of Fig.\ref{mag}(c).  In addition, we note that the maximum of this feature is suppressed for lower frequencies (we compare just two frequencies in the inset for clarity).  We speculate that this may originate from a small volume fraction of the crystal with sulfur vacancies that promote glassy magnetism or by domain effects associated with the long range order (i.e. a small, uncompensated moment due to canting).  However, we emphasize that we are not sure about the nature of this and the extent to which it involves the bulk. 

\begin{figure*}
\includegraphics[width=1.6\columnwidth]{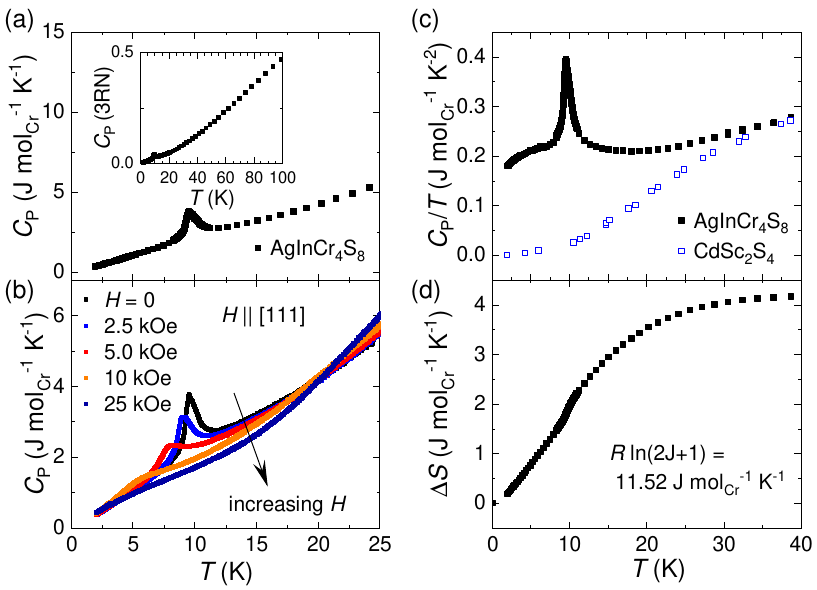} 
\caption{\label{entropy} (a) Specific heat capacity of \ce{AgInCr4S8} with inset showing a smooth evolution across larger temperature range in units normalized to the expected Dulong-Petite limit, (b) specific heat capacity with field applied along [111], (c) temperature-normalized specific heat capacity including data for \ce{CdSc2S4} extracted from Ref. \citenum{kitani2014} that is utilized as a phonon background to estimate the magnetic specific heat of \ce{AgInCr4S8}. (d) Entropy obtained by integrating the estimated magnetic contribution; a total entropy of the paramagnetic state is expected to be 11.52 J/mol/K per Cr$^{3+}$ having J = 3/2.}
\end{figure*}

The impact of an applied field on the temperature-dependent magnetization is shown in Fig. \ref{mag}(d).  As described above, a broad maximum in $M$ is observed near $T_{\mathrm N}$ for low fields.  With increasing $H$, the maximum is suppressed to lower temperature and is further broadened.  For intermediate fields surrounding 3 kOe, there is a small divergence between the ZFC and FC data.  This appears to correspond with the temperatures and fields where hysteresis is observed in the isothermal magnetization measurements, such as shown in Fig. \ref{mag}e for 1.9 K.  The $M(H)$ data are characterized by a rapid rise towards saturation with a subtle metamagnetic transition at low fields.  The corresponding feature manifests as non-linearity in $M(H)$ with a change in slope near 5 kOe at 1.9 K; the feature is rather broad, similar to the temperature-dependent data.  Although complete saturation of the moment does not occur until the highest applied field, it appears that roughly 25 kOe is sufficient to polarize the moments out of the antiferromagnetic state at 1.9 K.  With increasing $T$, the metamagnetic transition becomes more difficult to isolate and the saturation moment decreases, as shown in Fig. \ref{mag}(f).  Similarly, features are not observed in $M$($T$) for large fields as shown in Fig. \ref{mag}(d).

\begin{figure}
\includegraphics[width=0.9\columnwidth]{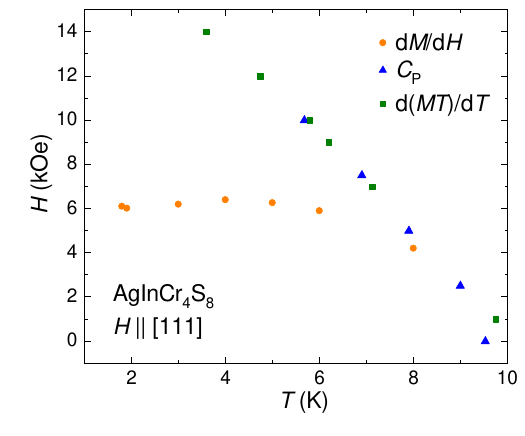} 
\caption{\label{phase} Magnetic phase diagram obtained from physical property data as indicated in the legend. The points selected from maximum in the d$M$/d$H$ curve are from magnetizing the sample after cooling in zero field; data for the d($MT$)/d$T$ were collected upon cooling in an applied field.}
\end{figure}

The specific heat capacity is shown in Fig. \ref{entropy}(a) in the region surrounding the Neel temperature with the inset showing a larger temperature range.  The anomaly in $C_P$ at the Neel temperature is broad and small but well-defined, suggesting this is associated with the onset of long-range magnetic order.  As shown in Fig. \ref{entropy}(b), the specific heat is continually suppressed by the application of a magnetic field.  This is consistent with the behavior observed for the magnetization in Fig. \ref{mag}(d).  To estimate the change in magnetic entropy at the Neel temperature, we integrated $C_P/T$ after subtracting an estimate of the lattice contribution.  The lattice contribution was estimated using the specific heat of \ce{CdSc2S4} taken from the literature;\cite{kitani2014} we note that Cd is between Ag and In on the periodic table.  The resulting entropy is plotted in Fig. \ref{entropy}(d).  It is seen that by 10\,K (just above $T_N$), the entropy reaches only $\approx$ 1/6 of the expected entropy for paramagnetic Cr$^{3+}$ spins.  By integrating up to roughly 40 K the entropy reaches roughly 1/3 of the expected entropy.  Based on this, we speculate that a large fraction of the magnetic entropy is consumed at high temperatures due to the presence of short-range correlations, as is typical in frustrated systems.

The original literature on \ce{AgInCr4S8} published in the 1970s contained very little data and the results were on polycrystalline samples.  The N\'eel temperature was reported as 14 K\cite{pinch1970some} and 17 K.\cite{plumier1971mise}  Plumier and Sougi reported a helimagnetic ground state for their powders in 1971,\cite{plumier1971mise} which is consistent with the results we will present below.  Later, in 1977, Plumier et al published additional results that showed anomalies in the specific heat capacity at 11.8 K and 42.5 K.\cite{plumier1977} The existence of two anomalies in the specific heat was claimed to be due to a change in magnetic structure with the helimagnetic ground state emerging from an existing ordered state not a paramagnetic state.  However, we did not observe any anomalies near 42 K in the specific heat for our single crystals, which we measured with intention using appropriate temperature spacings as well as the so-called large thermal pulse approach supported by Quantum Design.  Thus, we speculate that the specific heat anomaly occurring at 42 K in the 1977 manuscript may have been caused by an impurity in the polycrystalline sample.  Specifically, we suggest that this impurity was likely \ce{AgCrS2}, which is now known to have a strong magneto-structural transition near 42 K\cite{kawaji1989heat}; this is a first order transition that relieves magnetic frustration and long-range magnetic order is established.\cite{Damay2011}  Plumier suggested that their neutron diffraction results also supported the existence of order in \ce{AgInCr4S8} above the $T_{\mathrm N}$ $\approx$ 12 K, claiming magnetic order but with a correlation length of less than ten unit cells.\cite{plumier1977}  We are not able to assess this aspect directly due to our limited data.  Yet, given these specific heat results and our understanding of how growth conditions can impact samples, we believe the impact of impurities may have indeed been prominent, though we do not wish to speculate further due to the overall complexity of these breathing pyrochlore materials.  Exploring the nature of magnetic correlations above 10 K in \ce{AgInCr4S8} single crystals may prove to be interesting.  In addition, we note that the prior literature generally reported a higher transition temperature for their measurements on polycrystalline samples. It is not clear how the transition temperature was defined, though, and limited details of the synthesis are provided. While this makes it difficult to draw absolute conclusions through comparison with our crystals, the samples are evidently sensitive to the synthesis conditions.  Further study may reveal important links between synthesis conditions, defect formation, and the resulting magnetic behavior.

The magnetic transitions observed via the magnetization and specific data are summarized in the magnetic phase diagram illustrated in Fig. \ref{phase}. This illustrates that the transition from the high-temperature, correlated paramagnetic regime to the antiferromagnetic states is continuously suppressed by an applied magnetic field.  In the magnetic ground state, application of the magnetic field results in a weak metamagnetic transition that is illustrated in Fig. \ref{mag}(e).  This transition is illustrated by the orange circles, but we emphasize that the transition appears rather broad in the magnetization data.  As such, this phase diagram represents a qualitative average of the transition, but the phase border may be complex and in particular the transition could occur through an intermediate phase that is not shown in Fig. \ref{phase}.  Neutron diffraction in an applied field is required to understand the nature of the phases and whether the transition occurs through an intermediate phase.  We note that we selected to use the derivative d($MT$)/d$T$ for consistency with the specific heat for antiferromagnetic transitions.\cite{fisher1962relation}

\subsection{Magnetic Structure}

\begin{figure*}
\includegraphics[width=1.98\columnwidth]{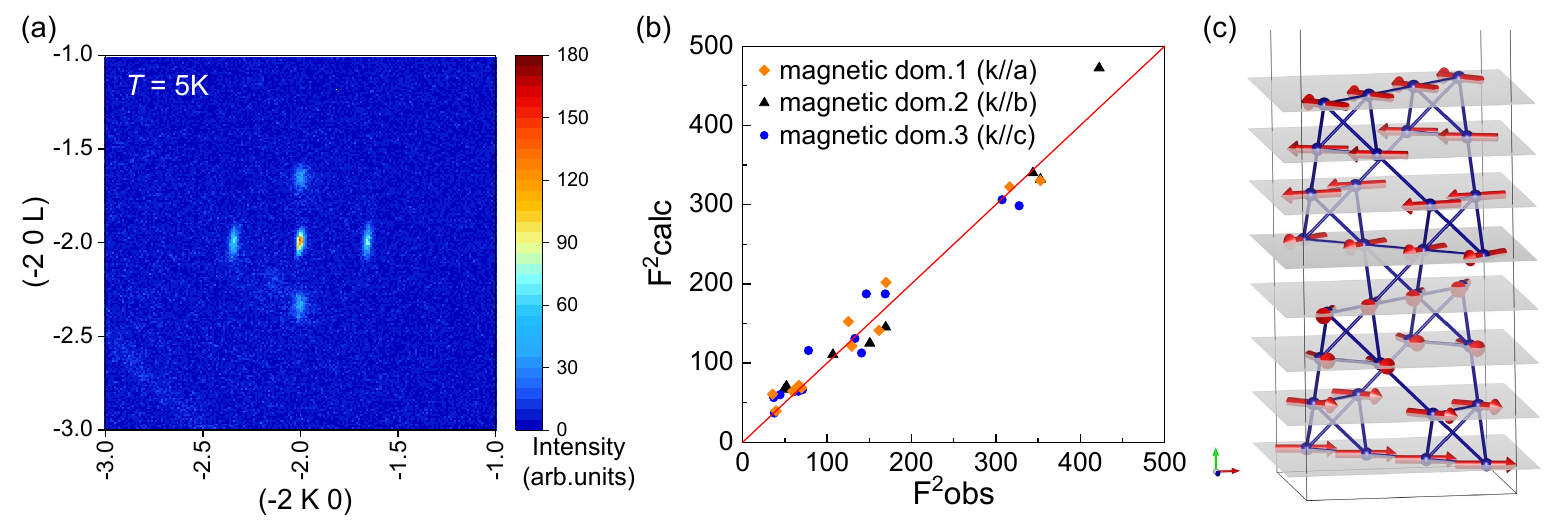} 
\caption{\label{neutron1} (a) Neutron diffraction intensity around the -2 -2 -2 nuclear Bragg peak showing four satellite peaks associated with the incommensurate magnetic order at $T$ = 5 K characterized by propagation vector $\textbf{\textit{k}}$ = (0, 0, $\pm$0.343). A weak intensity from a small secondary grain is observed to the lower left of the central peak. (b) Plot of observed vs calculated magnetic intensity for the fitted helical structure, which is shown in (c) for two lattice intervals along the $\textbf{\textit{k}}$-vector.  The spin structure consists of ferromagnetic layers of Cr moments with a modulation that follows the $\textbf{\textit{k}}$-vector.}
\end{figure*}

\begin{figure}
\includegraphics[width=0.85\columnwidth]{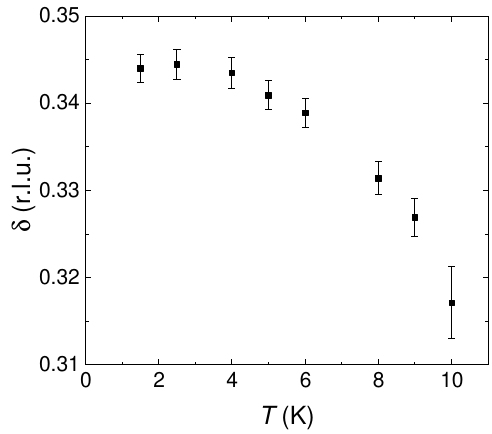} 
\caption{\label{neutron2} Incommensurate indexing $\delta$ as a function of temperature obtained via Gaussian fits to the magnetic (satellite) peaks near the nuclear (-1 -1 -1) and (-2 -2 -2) Bragg reflections.}
\end{figure}

Single crystal neutron diffraction data were collected at 5 K (below $T_{\mathrm N}$) to assess the nature of the ordered ground state. The measurements revealed incommensurate magnetic reflections corresponding to $\textbf{\textit{k}}$ = (0, 0, $\pm$0.343),  and its equivalents, ($\pm$0.343, 0, 0) and  (0, $\pm$0.343, 0).  Representative data are shown in Fig. \ref{neutron1} as a color-map of scattering intensity in the ($-2KL$) plane projection. The image displays the (-2, -2, -2) nuclear Bragg peak surrounded by four magnetic satellite peaks.  In addition to the four satellite peaks seen in Fig. \ref{neutron1}(a), satellites occur at approximately (-2-$\delta$,-2-2) and (-2+$\delta$,-2,-2). We also performed temperature-dependent neutron diffraction measurements around selected Bragg peaks to monitor how the incommensuration of the spin structure changes as a function of temperature.  The results shown in Fig. \ref{neutron2} demonstrate that $\delta$ increases upon cooling and essentially reaches a terminal value at base temperature (no feature or change in behavior was observed at the temperature of the onset of $\chi''$).  The incommensurate parameter also changes with temperature in other spinels, such as in the helical structure of \ce{ZnCr2Se4}.\cite{cameron2016magnetic}

Incommensurate magnetic ground states are also observed in \ce{CuGaCr4S8} and \ce{CuAlCr4S8} but the unit cell distorts to orthorhombic in these materials.  Interestingly, cubic symmetry is observed in the field-induced spin structure that yields a 1/2 magnetization plateau.\cite{Gen2025}  We did not detect any evidence for a structural distortion in \ce{AgInCr4S8}. However, we recognize that in related frustrated magnets, for example in \ce{ZnCr2Se4}, there is a very minor structural distortion at $T_{\mathrm N}$ that is beyond the resolution limits of neutron diffraction.\cite{hidaka2003structural,Hemberger2007}  Thus, it remains an open question whether the incommensurate ground state in \ce{AgInCr4S8} is coupled to a small structural distortion.  However, a distortion is not required within theories that have predicted such incommensurate states in cubic breathing pyrochlore lattices.\cite{ghosh2019breathing}

We now discuss our analysis of the magnetic structure obtained by refining the $T$ = 5 K dataset.  By considering the space group $F4\bar{3}m$ and the incommensurate propagation vector (0, 0, 0.343), the magnetic representations of the little group $G_k$ can be decomposed into four one-dimensional irreducible representations. The Cr atom site is split into two orbits, meaning that the magnetic order can potentially be composed from two independent sublattices. The irreducible representation in Kovalev notation and the associated basis vectors for each Cr orbit is presented in Table \ref{tab_irreps} of the Appendix. We found that the experimental data are best described by the irreducible representations $\Gamma_3$ and $\Gamma_4$ which allow ferromagnetic coupling between nearest-neighbors Cr atoms within the plane perpendicular to the $\textbf{\textit{k}}$ vector. To define a helical magnetic structure, as previously considered for this system and similar materials, \cite{plumier1971mise,plumier1966neutron,hastings1968magnetic} a combination of the two representations $\Gamma_3 \oplus \Gamma_4$ is required. 

Given the uncertainty arising from data limitations and the potential complexity of the underlying spin model, we restricted the present analysis to a simple helical structure.  More complex arrangements may exists, such as non-coplanar spin structures or multi-$\textbf{\textit{k}}$ orderings that are difficult to differentiate based on diffraction alone. We note that the possible presence of multi-\textbf{\textit{k}} magnetic structures could, in principle, be investigated through measurements of the magnetic excitation spectrum using inelastic neutron scattering, provided that sufficiently large single crystals become available. The helical magnetic structure model proposed to describe our data consists of ferromagnetic layers of Cr moments that rotate as they translate along the $\textbf{\textit{k}}$-direction. The Cr atoms belonging to different orbits are related by a phase factor of exp(2$\pi\textbf{\textit{k}}\cdot$ r$_{ij}$), where r$_{ij}$ is the translation vector between the relative atomic positions. 

The parent space group is non-centrosymmetric and $\textbf{\textit{k}}$ is not equivalent to $\textbf{\textit{-k}}$. Therefore, a proper symmetry analysis requires consideration of the extended little group $G_{k,-k}$.  Under these conditions, the most consistent and rigorous description of the magnetic ordering is provided by the superspace group formalism, which naturally incorporates the symmetry constraints imposed by incommensurability and the lack of inversion. Using the program ISODISTORT \cite{campbell2006isodisplace,isodistort} we determined that there are three physical irreducible representations of $G_{k,-k}$ ($\Gamma_{mag}$ = $mDT_1$ +  $mDT_2$ + $mDT_3\cdot DT_4$), with all four Cr moments contained within a single orbit. The helimagnetic model with ferromagnetic layers, as described above, corresponds to the $mDT_3\cdot DT_4$ irreducible representation and the superspace group $F222.1’(0, 0, g)00ss$. 

Within the superspace formalism, a helical magnetic modulation is naturally described as superposition of sinusoidal modulations along orthogonal directions, typically represented by sine and cosine functions of the internal coordinate (x$_4$), which parameterize the phase of the modulation along the propagation vector $\textbf{\textit{k}}$. In the superspace formalism, this additional coordinate represents the internal phase of the incommensurate modulation, allowing the magnetic structure to be described as periodic in (3+1)-dimensional space.\cite{perez2012magnetic} Although the space-group symmetry allows moment components along all three crystallographic directions, we constrained the modulation of Cr moments to lie perpendicular to the $\textbf{\textit{k}}$-vector. Additionally, we imposed equal sine and cosine amplitudes to enforce a simple circular helix. The data for each of the three symmetry-related domains arising from the cubic crystal lattice are fit assuming an equal domain population in the crystal. This model yields an agreement factor of R$_f$ = 7\% on 32 distinct magnetic reflections. The refined moment value is m(Cr) = 2.2 $\pm$ 0.1 $\mu_B$/Cr.  A schematic representation of the magnetic structure is shown in Fig. \ref{neutron1}(c), while the observed versus calculated intensity comparison is presented in Fig. \ref{neutron1}(b). The full superspace description of the magnetic structure is given in Table \ref{tab_magparams}. Diffraction data will be available at a DOI upon publication.\cite{doi}

Simulated powder diffraction indicates that the strongest magnetic reflection occurs at (0, 0, 0.343), which was not accessible in the present single-crystal diffraction experiment. Further studies of the temperature and applied magnetic-field evolution of the magnetic structure would benefit from access to this low-$Q$ magnetic reflection.  Such measurements would require experimental configurations capable of accessing an intermediate $Q$ range that lies beyond the typical coverage of small-angle neutron scattering and below that of conventional single-crystal diffraction experiments.

The theoretically predicted magnetic ground state for the exchange parameters J, J' $<$ 0 and J$_{3a}$ $>$ J$_{3b}$  $>$ 0 gives rise to a spin spiral modulation consistent with the observed propagation vector ($\delta$, 0, 0); see Ref. \citenum{ghosh2019breathing} for a detailed discussion of the theoretical framework and the corresponding interaction bonds for these parameters. While in the present spin model all Cr atoms belong to a single symmetry orbit, in alternative symmetry descriptions where the Cr sites split into symmetry-inequivalent positions, the magnetic structure could instead support slightly out-of-phase spiral modulations. Although such a pitch mismatch was not considered in our magnetic structure model, it is expected to be on the order of a few degrees and would result in subtle variations in the intensity of magnetic Bragg peaks. The degree of this mismatch is thought to be sensitive to the extent of breathing distortion in the Cr sublattice \cite{ghosh2019breathing}, suggesting that future spectroscopic studies that can probe unusual interaction profiles induced by this distortion can further elucidate the underlying microstructure.

\section{Summary}
This study examines the magnetism in single crystalline specimens of a breathing pyrochlore spinel, a family of materials where single crystals are rare.  Crystals grown in an excess of sulfur were found to display physical properties and Bragg diffraction consistent with long-range magnetic order below $T_{\mathrm N}$=9.6(2) K.  While not discussed in detail, it is important to emphasize that crystals grown without excess sulfur displayed properties that are reminiscent of spin freezing as opposed to long-range order, and thus small changes in chemistry can drive large changes in the magnetism and therefore crystal growth needs to be approached with the utmost care.

Single crystal neutron and x-ray diffraction data confirmed the non-centrosymmetric symmetry associated with Ag/In ordering that yields the breathing pyrochlore magnetic sublattice of Cr ions.  This architecture is known to promote geometric frustration and indeed our specific heat capacity data suggest that short-range correlations persist at many multiples of $T_{\mathrm N}$. Magnetization measurements reveal a weak field-induced transition below 10\,kOe and saturation near the expected 3$\mu_B$ at roughly 25 kOe.

The simplest magnetic model that describes the incommensurate structure probed by neutron diffraction is a helical structure described by the superspace group $F222.1'(0,0,g)00ss$.  In this model, all Cr moments are aligned ferromagnetically within individual layers, which rotate uniformly around the propagation vector $\textbf{\textit{k}}$ = (0, 0, $\delta$), forming a helical magnetic structure. The refined magnetic moment is approximately 2.2 $\mu_B$/Cr atom. Given uncertainty due to data limitations and the potential complexity of the spin structure, we here limit the presented results to the simple helical structure with the caveat that structures with greater complexity, such as non-coplanar spin structures could perhaps be more suitable.  Investigation of the inelastic spectra could help to further codify the nature of the ground state, though the crystal size obtained herein is challenging for such experiments on current neutron scattering instrumentation.

\begin{acknowledgments}
 This research was supported by the U.S. Department of Energy, Office of Science, Basic Energy Sciences, Materials Science and Engineering Division. A portion of this research used resources at the High Flux Isotope Reactor, a DOE Office of Science User Facility operated by the Oak Ridge National Laboratory. The beam time was allocated to WAND$^2$ on proposal number IPTS-29635.1. 
\end{acknowledgments}


\providecommand{\noopsort}[1]{}\providecommand{\singleletter}[1]{#1}%

\pagebreak

\section{Appendix}
\appendix

\clearpage
\onecolumngrid
 \section{Tables associated with the determinations of the magnetic structure from neutron diffraction data}

\renewcommand{\thefigure}{A\arabic{figure}}
\renewcommand{\thetable}{A\Roman{table}}
\renewcommand{\thesection}{S\Roman{section}}
\setcounter{figure}{0}
\setcounter{table}{0}

\begin{table*}[]
\centering
\caption{Irreducible representation and basis vectors of the space group $F\text{-}43m$ and the incommensurate propagation vector $\mathbf{k} = (0, 0, 0.34)$. The Cr atom site is split into two orbits Cr1 and Cr2.}
\label{tab_irreps}
\begin{tabular}{|c|c|c|c|c|}
\hline
Irrep & 
\begin{tabular}[c]{@{}l@{}}Cr1\_1 (x, y, z)\\ (0.628, 0.128, 0.628)\end{tabular} & 
\begin{tabular}[c]{@{}l@{}}Cr1\_2 ($-x$+1, y, $-z$+1)\\ (0.371, 0.128, 0.371)\end{tabular} & 
\begin{tabular}[c]{@{}l@{}}Cr2\_1 (x, y, z)\\ (0.371, 0.871, 0.628)\end{tabular} & 
\begin{tabular}[c]{@{}l@{}}Cr2\_2 ($-x$+1, y, $-z$+1)\\ (0.628, 0.871, 0.371)\end{tabular} \\ 
\hline
$\Gamma_1$ & 
(1 \;\; -1 \;\; 0) & 
(-1 \;\; 1 \;\; 0) & 
(1 \;\; 1 \;\; 0) & 
(-1 \;\; -1 \;\; 0) \\ 
\hline
$\Gamma_2$ & 
\begin{tabular}[c]{@{}l@{}}(1 \;\; 1 \;\; 0)\\ (0 \;\; 0 \;\; 1)\end{tabular} & 
\begin{tabular}[c]{@{}l@{}}(-1 \;\; -1 \;\; 0)\\ (0 \;\; 0 \;\; 1)\end{tabular} & 
\begin{tabular}[c]{@{}l@{}}(1 \;\; -1 \;\; 0)\\ (0 \;\; 0 \;\; 1)\end{tabular} & 
\begin{tabular}[c]{@{}l@{}}(-1 \;\; 1 \;\; 0)\\ (0 \;\; 0 \;\; 1)\end{tabular} \\ 
\hline
$\Gamma_3$ & 
(1 \;\; -1 \;\; 0) & 
(1 \;\; -1 \;\; 0) & 
\begin{tabular}[c]{@{}l@{}}(1 \;\; -1 \;\; 0)\\ (0 \;\; 0 \;\; 1)\end{tabular} & 
\begin{tabular}[c]{@{}l@{}}(1 \;\; -1 \;\; 0)\\ (0 \;\; 0 \;\; -1)\end{tabular} \\ 
\hline
$\Gamma_4$ & 
\begin{tabular}[c]{@{}l@{}}(1 \;\; 1 \;\; 0)\\ (0 \;\; 0 \;\; 1)\end{tabular} & 
\begin{tabular}[c]{@{}l@{}}(1 \;\; 1 \;\; 0)\\ (0 \;\; 0 \;\; -1)\end{tabular} & 
(1 \;\; 1 \;\; 0) & 
(1 \;\; 1 \;\; 0) \\ 
\hline
\end{tabular}
\end{table*}

\begin{table*}[]
\centering
\caption{Magnetic structure parameters for the modulated phase.}
\label{tab_magparams}
\begin{tabular}{|l|l|}
\hline
Parent space group & F$\text{-}43m$ \\ \hline
Propagation vector & (0, 0, 0.343) \\ \hline
Magnetic Superspace group & $F222.1'(0,0,g)00ss$ \\ \hline
Magnetic point group & 222.1' \\ \hline
Primary irrep(s) & $mDT_3\cdot DT_4$ \\ \hline
\begin{tabular}[c]{@{}l@{}}Setting of the subgroup \\ with respect to the parent space group\end{tabular} & 
\begin{tabular}[c]{@{}l@{}}basis = (1, 0, 0, 0), (0, 0, 1, 0), (0, -1, 0, 0), (0, 0, 0, 1) \\ origin = (0, 0, 0, 0)\end{tabular} \\ \hline
Unit cell parameters at $T$ = 5 K & a = b = c = 10.1693 $\AA$; $\alpha$ = $\beta$ = $\gamma$ = 90$^{\circ}$ \\ \hline
Symmetry operations & 
\begin{tabular}[c]{@{}l@{}}
1. $x_1$, $x_2$, $x_3$, $x_4$, +1 \\
2. $x_1$, $-x_2$, $-x_3$, $-x_4$, +1 \\
3. $-x_1$, $x_2$, $-x_3$, $-x_4$ + 1/2, +1 \\
4. $-x_1$, $-x_2$, $x_3$, $x_4$ + 1/2, +1
\end{tabular} \\ \hline
\begin{tabular}[c]{@{}l@{}}Symmetry centering \\ operations\end{tabular} & 
\begin{tabular}[c]{@{}l@{}}
1. $x_1$, $x_2$, $x_3$, $x_4$, +1 \\
2. $x_1$, $x_2$, $x_3$, $x_4$ + 1/2, -1 \\
3. $x_1$, $x_2$ + 1/2, $x_3$ + 1/2, $x_4$, +1 \\
4. $x_1$, $x_2$ + 1/2, $x_3$ + 1/2, $x_4$ + 1/2, -1 \\
5. $x_1$ + 1/2, $x_2$, $x_3$ + 1/2, $x_4$, +1 \\
6. $x_1$ + 1/2, $x_2$, $x_3$ + 1/2, $x_4$ + 1/2, -1 \\
7. $x_1$ + 1/2, $x_2$ + 1/2, $x_3$, $x_4$, +1 \\
8. $x_1$ + 1/2, $x_2$ + 1/2, $x_3$, $x_4$ + 1/2, -1
\end{tabular} \\ \hline
Positions of magnetic (Cr) atoms & (0.622, 0.123, 0.377), $(m_x, m_y, m_z)$ \\ \hline
\begin{tabular}[c]{@{}l@{}}Magnetic moment components ($\mu_B$) \\ and total magnitude\end{tabular} & 
\begin{tabular}[c]{@{}l@{}}
$M_x$ (cos) = 2.25 $\mu_B$ \\
$M_y$ (sin) = 2.25 $\mu_B$ \\
$M_z$ = 0 $\mu_B$ \\
$m_\mathrm{tot}$(Cr) = 2.2(1) $\mu_B$
\end{tabular} \\ \hline
\begin{tabular}[c]{@{}l@{}}Non-magnetic atoms (site multiplicity) \\ and atomic coordinates\end{tabular} & 
\begin{tabular}[c]{@{}l@{}}
In (4c) \hspace{1em} 0.25, 0.25, 0.25 \\
Ag (4b) \hspace{1em} 0.00, 0.00, 0.50 \\
S1 (16k) \hspace{0.5em} 0.3918, 0.3918, 0.1081 \\
S2 (16k) \hspace{0.5em} 0.8638, 0.3638, 0.1360
\end{tabular} \\ \hline
\end{tabular}
\end{table*}

\end{document}